\documentclass[journal,transmag]{IEEEtran}
\ifCLASSINFOpdf
  \usepackage[pdftex]{graphicx}
\else
\fi
\usepackage[cmex10]{amsmath}
\usepackage{amsfonts}
\begin{document}
%
\title{Power Spectral Density of Magnetization Dynamics Driven by a Jump-Noise Process}


\author{\IEEEauthorblockN{A. Lee\IEEEauthorrefmark{1},
G. Bertotti\IEEEauthorrefmark{2},
C. Serpico\IEEEauthorrefmark{3}, and
I. Mayergoyz\IEEEauthorrefmark{1}}
\IEEEauthorblockA{\IEEEauthorrefmark{1}University of Maryland, College Park, MD 20740 USA}
\IEEEauthorblockA{\IEEEauthorrefmark{2}INRIM, Torino, Italy}
\IEEEauthorblockA{\IEEEauthorrefmark{3}University of Naples Federico II, Naples, Italy}%
\thanks{Corresponding author: I. Mayergoyz (email: isaak@umd.edu)}}

%



\IEEEtitleabstractindextext{%
\begin{abstract}
Random magnetization dynamics driven by a jump-noise process is reduced to stochastic magnetic energy dynamics on specific graphs using an averaging technique.  An approach to analyzing stochastic energy dynamics on graphs is presented and applied to the calculation of power spectral density of random magnetization dynamics.  An eigenvalue technique for computing the power spectral density under specific cases is also presented and illustrated by numerical results.
\end{abstract}

}

\maketitle

\IEEEdisplaynontitleabstractindextext

%
\IEEEpeerreviewmaketitle

\section{Introduction}
%
%
%
%
\IEEEPARstart{I}{n} recent years, stochastic magnetization dynamics has been the focus of considerable research due to its scientific importance and promising technological applications.  It has been recently proposed [1,2] that a jump-noise process can be used to describe thermal bath effects on magnetization dynamics.  A distinct advantage of this approach is that both the damping and fluctuation effects emerge from the nature of the jump-noise process.  This jump-noise process term in the equation for magnetization dynamics is usually small in comparison with the precessional term. This term leads to damping and fluctuation effects that occur on a much longer time-scale than magnetic precessions.  For this reason, an averaging technique can be applied to the random magnetization dynamics to reduce it to stochastic magnetic energy dynamics on graphs.  In this paper, an approach to the analysis of stochastic energy dynamics on graphs is presented and applied to the calculation of power spectral density of random magnetization dynamics.  By using the differential equation for transition probability density, formulas for the autocovariance function and the power spectral density are derived and illustrated by numerical computations.

\section{Technical Discussion}
Random magnetization dynamics driven by a jump-noise process is governed by the equation,
\begin{equation}
\frac{d\mathbf{M}}{dt}=-\gamma (\mathbf{M}\times\mathbf{H}_{eff})+\mathbf{T}_r(t),
\label{eq1}
\end{equation}
where $\mathbf{M}$ is the magnetization vector, $\gamma$ is the gyromagnetic ratio, $\mathbf{H}_{eff}$ is an effective magnetic field, and $\mathbf{T}_r(t)$ is a jump-noise torque that describes the effects of the thermal bath.  The random process $\mathbf{T}_r(t)$ is described by the formula
\begin{equation}
\mathbf{T}_r(t)=\sum_{i=1}\mathbf{m}_i\delta(t-t_i),
\label{eq2}
\end{equation}
where $\mathbf{m}_i$ are random jumps in magnetization occurring at random time instances $t_i$.  The statistics of the random jumps $\mathbf{m}_i$ and random times $t_i$ can be fully specified (see [1,2]) by introducing the transition probability rate $S(\mathbf{M}_i,\mathbf{M}_{i+1})$ with $\mathbf{M}_i=\mathbf{M}(t_i^-)$ and $\mathbf{M}_{i+1}=\mathbf{M}(t_i^+)=\mathbf{M}_i+\mathbf{m}_i$ where $\mathbf{M}(t_i^-)$ and $\mathbf{M}(t_i^+)$ are the magnetization immediately before and after a jump. Using the transition probability rate $S(\mathbf{M}, \mathbf{M}^\prime)$, the statistics of random times $t_i$ and random jumps $\mathbf{m}_i$ are defined by the formulas
\begin{equation}
Probability(t_{i+1} - t_i > \tau) = \exp \left[ - \int_{t_i}^{t_{i+1}} \lambda (\mathbf{M}(t)) dt \right],
\label{eq3}
\end{equation} 
\begin{equation}
\lambda(\mathbf{M}(t)) = \oint_{\Sigma} S(\mathbf{M}(t), \mathbf{M}^\prime ) d \Sigma_{\mathbf{M}^\prime},
\label{eq4}
\end{equation}
\begin{equation}
\chi (\mathbf{m}_i | \mathbf{M}_i) = \frac{S(\mathbf{M}_i, \mathbf{M}_i + \mathbf{m}_i)}{\lambda (\mathbf{M}_i)}.
\label{eq5}
\end{equation}
Here: $\lambda (\mathbf{M})$ is the scattering rate, $\chi(\mathbf{m}_i |\mathbf{M}_i)$ is the probability density for magnetization jumps $\mathbf{m}_i$ at time $t_i$ from the states $\mathbf{M}(t)=\mathbf{M}_i$, and $\Sigma$ is the sphere $|\mathbf{M}^\prime | = M_s = const$.

The following formula has been suggested in [1] for transition probability rate
\begin{equation}
S(\mathbf{M},\mathbf{M}')=Ae^{-\frac{|\mathbf{M}-\mathbf{M}'|^2}{2\sigma^2}}e^{\frac{g(\mathbf{M})-g(\mathbf{M}')}{2kT}},
\label{eq6}
\end{equation}
where A is an empirically derived constant characterizing the strength of the jump-noise process $\mathbf{T}_r(t)$.

It is important to note that the stochastic magnetization dynamics (\ref{eq1}) can also be equivalently described in terms of a Kolmogorov-Fokker-Planck equation for transition probability density $w(\mathbf{M},t)$:
\begin{align}
\frac{\partial w}{\partial t}&=-\gamma\nabla_\Sigma\cdot[(\mathbf{M}\times\nabla_\Sigma g)w] \\
\nonumber & +\int_\Sigma [S(\mathbf{M}',\mathbf{M})w(\mathbf{M}')-S(\mathbf{M},\mathbf{M}')w(\mathbf{M})]d\Sigma_{\mathbf{M}'}
\label{eq77}
\end{align}
where $g$ is the micromagnetic energy.

The last equation is linear and deterministic with respect to $w$. This is the main advantage of equation (7) in comparison with equation (1), which is nonlinear stochastic differential equation.

Usually, damping and fluctuations caused by the thermal bath occur on a much longer time-scale than the fast motion of magnetization precessions.  If one's interest lies in analyzing only the thermal effects, then the magnetization dynamics described by equation (\ref{eq1}) can be averaged over precessional trajectories, which are uniquely defined by their micromagnetic energy $g$.  This averaging leads to a description of stochastic dynamics for $g$ that is defined on specific energy graphs.  These graphs reflect the energy landscapes on $\Sigma$ of magnetic particles.

By using equation (7) and averaging both sides over precessional trajectories in the manner discussed in [6], it can be shown that in terms of the transition probability density $\rho(g,t)$, the stochastic energy dynamics is described by the following equation
\begin{align}
\frac{d\rho_k(g,t)}{dt}& =\sum_n\int_{L_n} [K_{n,k}(g',g)\rho_n(g',t) \\
\nonumber & -K_{k,n}(g,g')\rho_k(g,t)]dg', (k=1,2,...N),
\label{eq8}
\end{align}
where $L_n$ is an edge of the graph corresponding to the region $R_n$ of the sphere $\Sigma$ with the property that there exists only one precessional trajectory $C_n(g)$ corresponding to energy $g$, $\rho_n(g,t)$ is the probability density on edge $L_n$, and function $K_{n,k}(g',g)$ is related to $S(\mathbf{M}',\mathbf{M})$ by the formula
\begin{align}
K_{n,k}&(g',g)= \\
\nonumber &\frac{1}{\tau_n(g)}\oint_{C_k(g)}\oint_{C_n(g')}\frac{S(\mathbf{M}',\mathbf{M})}{|\nabla_\Sigma g(\mathbf{M}')||\nabla_\Sigma g(\mathbf{M})|}dl_{\mathbf{M}'}dl_{\mathbf{M}},
\label{eq9}
\end{align}
where
\begin{equation}
\tau_n(g)=\oint_{C_n(g)}\frac{dl_{\mathbf{M}}}{|\nabla_\Sigma g(\mathbf{M})|}.
\label{eq10}
\end{equation}
In equation (8), the summation is performed over all N edges of the graph.

The effect of spin-torque can be included by modifying equation (8) as follows
\begin{align}
\frac{\partial}{\partial t}\rho_k(g,t)=&\frac{\partial}{\partial g}[\Phi_k(g)\rho_k(g,t)]
\nonumber \\ &+\sum_n\int_{L_n}[K_{n,k}(g',g)\rho_n(g',t)
\nonumber \\ & \qquad \qquad -K_{k,n}(g,g')\rho_k(g,t)]dg',
\label{eq11}
\end{align}
where $\Phi_k(g)$ is a function that describes the effect of spin-torque as discussed in [8].

Now, we proceed to the discussion of computation of the power spectral density of random micromagnetic energy $g$. In control theory, the power spectral density is computed for linear time-invariant systems. For such systems, the power spectral density of the output is equal to the power spectral density of the input multiplied by the squared magnitude of the transfer function. However, the magnetization dynamics described by the stochastic differential equation (1) is nonlinear. Nevertheless, it turns out that the power spectral density can be computed by using linear techniques. This can be done by exploiting the linearity of equations (7) and (8) for transition probability densities $w$ and $\rho$, respectively. This approach is used in our subsequent discussion. 

The power spectral density is defined by the formula
\begin{equation}
\hat{S}_f(\omega)=\int_{-\infty}^{\infty}\hat{C}_f(\tau)e^{-j\omega\tau}d\tau,
\label{eq12}
\end{equation}
where $\hat{C}_f(\tau)$ is the autocorrelation function.  This autocorrelation function is given by the formula
\begin{align}
\hat{C}_f(\tau) =&\int_L\int_L f(g)f(g')\rho(g,t_0;g',t_0-\tau)dg'dg
\nonumber \\&-\int_L\int_L f(g)f(g')\rho^{eq}(g)\rho^{eq}(g')dg'dg.
\label{eq16}
\end{align}
where $\rho^{eq}(g)$ is the equilibrium probability density.

For a stationary Markovian process, the joint probability density can be expressed by the formula
\begin{equation}
\rho(g,t_0;g',t_0-\tau)=\rho(g,\tau|g',0)\rho^{eq}(g').
\label{eq17}
\end{equation}
Using the formula (\ref{eq17}), equation (\ref{eq16}) can be written as
\begin{align}
\begin{split}
\hat{C}_f(\tau) &=\\
\int_L f(g)\bigg[\int_L f(g')[\rho(g,\tau|g',0)&-\rho^{eq}(g)]\rho^{eq}(g')dg'\bigg]dg.
\end{split}
\label{eq18}
\end{align}
The expression within the brackets in equation (\ref{eq18}) is
\begin{equation}
\psi_f(g,\tau)=\int_L f(g')[\rho(g,\tau|g',0)-\rho^{eq}(g)]\rho^{eq}(g')dg'.
\label{eq19}
\end{equation}
Therefore, formula (\ref{eq16}) can be expressed as
\begin{equation}
\hat{C}_f(\tau)=\int_L f(g)\psi_f(g,\tau)dg.
\label{eq20}
\end{equation}
Using equation (\ref{eq20}), the power spectral density can be written as follows
\begin{equation}
\hat{S}_f(\omega)=2Re\left\{\int_L f(g)\left[\int_{0}^{\infty}\psi_f(g,\tau)e^{-j\omega\tau}d\tau\right]dg\right\}.
\label{eq21}
\end{equation}
The expression in the inner bracket of formula (\ref{eq21}) can be seen as the following Fourier Transform 
\begin{equation}
\Psi_f(g,\omega)=\int_{0}^{\infty}\psi_f(g,\tau)e^{-j\omega\tau}d\tau.
\label{eq22}
\end{equation}
Therefore, the power spectral density can be written as
\begin{equation}
\hat{S}_f(\omega)=2Re\left\{\int_L f(g)\Psi_f(g,\omega)dg\right\}.
\label{eq23}
\end{equation}

When we are interested in the spectral density of $g$, equation (\ref{eq20}) is reduced to
\begin{equation}
\hat{C}_g(\tau)=\int_L g\psi_g(g,\tau)dg,
\label{eq24}
\end{equation}
where
\begin{equation}
\psi_g(g,\tau)=\int_L g'[\rho(g,\tau|g',0)-\rho^{eq}(g)]\rho^{eq}(g')dg'.
\label{eq25}
\end{equation}
It is clear now that equation (\ref{eq11}) can be used to compute the power spectral density of $g$.

In formula (\ref{eq11}), $\rho(g,t)$ is the simplified notation for $\rho(g,t_0+\tau|g_0,t_0)$. For a stationary process,
\begin{equation}
\rho(g,t_0+\tau|g_0,t_0)=\rho(g,\tau|g_0,0).
\label{eq26}
\end{equation}
Using equation (\ref{eq26}), equation (\ref{eq11}) can be written as
\begin{align}
\frac{\partial}{\partial \tau}&\rho_k(g,\tau|g_0,0)=\frac{\partial}{\partial g}[\Phi_k(g)\rho_k(g,\tau|g_0,0)]
\nonumber \\ &+\sum_n\int_{L_n}[K_{n,k}(g',g)\rho_n(g',\tau|g_0,0)
\nonumber \\ & \qquad \qquad -K_{k,n}(g,g')\rho_k(g,\tau|g_0,0)]dg'.
\label{eq27}
\end{align}
At $t=0$, the following initial condition is valid:
\begin{equation}
\rho(g_0,0;g,\tau)|_{\tau=0}=\delta(g-g_0).
\label{eq28}
\end{equation}
At equilibrium, the stationary probability distribution $\rho^{eq}$ satisfies the equation
\begin{align}
\frac{\partial}{\partial g}[\Phi_k(g)\rho^{eq}_k(g)]+&\sum_n\int_{L_n}[K_{n,k}(g',g)\rho^{eq}_n(g')
\nonumber \\ &-K_{k,n}(g,g')\rho^{eq}_k(g)]dg'=0.
\label{eq29}
\end{align}
Taking the difference between equations (\ref{eq27}) and (\ref{eq29}) leads to the formula
\begin{align}
\frac{\partial}{\partial \tau}&[\rho_k(g,\tau|g_0,0)-\rho_k^{eq}(g)]
\nonumber \\ &=\frac{\partial}{\partial g}[\Phi_k(g)[\rho_k(g,\tau|g_0,0)-\rho_k^{eq}(g)]]
\nonumber \\ &+\sum_n\int_{L_n}[K_{n,k}(g',g)[\rho_n(g,\tau|g_0,0)-\rho_n^{eq}(g)]
\nonumber \\ &\qquad\qquad-K_{k,n}(g,g')[\rho_k(g,\tau|g_0,0)-\rho_k^{eq}(g)]]dg'.
\label{eq30}
\end{align}
Now, we introduce the function
\begin{equation}
\zeta_k (g,\tau|g_0,0)\equiv\rho_k(g,\tau|g_0,0)-\rho_k^{eq}(g).
\label{eq31}
\end{equation}
The initial condition for this function is:
\begin{equation}
\zeta_k(g,\tau|g_0,0)|_{\tau=0}=\delta(g-g_0)-\rho_k^{eq}(g).
\label{eq32}
\end{equation}
Using formula (\ref{eq31}), equation (\ref{eq30}) can be written in the form
\begin{align}
\frac{\partial}{\partial \tau}&\zeta_k(g,\tau|g_0,0)=\frac{\partial}{\partial g}[\Phi_k(g)\zeta_k(g,\tau|g_0,0)]
\nonumber \\ &+\sum_n\int_{L_n}[K_{n,k}(g',g)\zeta_n(g',\tau|g_0,0)
\nonumber \\ & \qquad \qquad -K_{k,n}(g,g')\zeta_k(g,\tau|g_0,0)]dg'.
\label{eq33}
\end{align}
By solving for $\zeta$, function $\psi_g(g, \tau)$ in equation (\ref{eq25}) can also be found:
\begin{equation}
\psi_k(g,\tau)\equiv\int_L g'\rho_k^{eq}(g')\zeta_k(g,\tau|g',0)dg'.
\label{eq34}
\end{equation}
Using formula (\ref{eq34}), equation (\ref{eq33}) can be transformed as follows:
\begin{align}
\frac{\partial}{\partial \tau}&\psi_k(g,\tau|g_0,0)=\frac{\partial}{\partial g}[\Phi_k(g)\psi_k(g,\tau|g_0,0)]
\nonumber \\ &+\sum_n\int_{L_n}[K_{n,k}(g',g)\psi_n(g',\tau|g_0,0)
\nonumber \\ & \qquad \qquad -K_{k,n}(g,g')\psi_k(g,\tau|g_0,0)]dg'.
\label{eq35}
\end{align}
It follows from equation (\ref{eq32}) and (\ref{eq34}) that the initial condition for $\psi_k$ is: 
\begin{equation}
\psi_k(g,\tau|g_0,0)|_{\tau=0}=\rho_k^{eq}[g-\langle g \rangle].
\label{eq36}
\end{equation}
Applying the one-sided Fourier transform in (\ref{eq22}) to equation (\ref{eq35}) results in:
\begin{align}
j\omega&\Psi_k(g,\omega|g_0)-\psi(g,\tau|g_0,0)|_{\tau=0}=
\nonumber \\ & \frac{\partial}{\partial g}[\Phi_k(g)\Psi_k(g,\omega|g_0)]
\nonumber \\ &+\sum_n\int_{L_n}[K_{n,k}(g',g)\Psi_n(g',\omega|g_0)
\nonumber \\ & \qquad \qquad -K_{k,n}(g,g')\Psi_k(g,\omega|g_0)]dg'.
\label{eq37}
\end{align}
This equation can now be used to compute $\Psi_k$.  The power spectral density can then be found as:
\begin{equation}
\hat{S}_g(\omega)=2Re\left\{\int_L g\Psi(g,\omega)dg\right\}.
\label{eq38}
\end{equation}

In the case of no applied spin-torque, equation (\ref{eq37}) can be simplified to
\begin{align}
j\omega&\Psi_k(g,\omega|g_0)-\psi_k(g,\tau|g_0,0)|_{\tau=0}
\nonumber \\ &=\sum_n\int_{L_n}[K_{n,k}(g',g)\Psi_n(g',\omega|g_0)
\nonumber \\ & \qquad \qquad -K_{k,n}(g,g')\Psi_k(g,\omega|g_0)]dg'.
\label{eq39}
\end{align}
By numerically solving equation (34) or the last equation, functions $\Psi_k (g,\omega)$ can be found and then they can be used in equation (35) to compute the power spectral density $S_g(\omega)$. In the case of no applied spin-torque, a special technique of solving equation (36) can be useful. The right-hand side of this equation has the form of the collision integral and, for this reason, it can be solved using an eigenvalue approach:
\begin{equation}
\hat{K}\phi_i=\lambda_i\phi_i
\label{eq40}
\end{equation}
where $\hat{K}$ is the collision integral operator in (\ref{eq39}).

Function $\Psi_k(g,\omega)$ can be decomposed with respect to eigenfunctions of $\hat{K}$:
\begin{equation}
\Psi_k(g,\omega)=\sum_i a_i(\omega) \phi_i(g).
\label{eq41}
\end{equation}

Similarly,
\begin{equation}
\psi_k(g,\tau|g_0,0)|_{\tau=0}=\sum_i b_i \phi_i(g),
\label{eq42}
\end{equation}
where
\begin{equation}
b_i=\langle \rho_k^{eq}[g-\langle g \rangle], \phi_i(g)\rangle.
\label{eq43}
\end{equation}

Now, equation (\ref{eq39}) can be reduced to
\begin{equation}
j\omega a_i(\omega)\phi_i(g)-b_i \phi_i(g)=\lambda_i a_i(\omega)\phi_i(g).
\label{eq44}
\end{equation}
Consequently,
\begin{equation}
a_i=\frac{b_i}{j\omega-\lambda_i}.
\label{eq45}
\end{equation}
Therefore, we have
\begin{equation}
\Psi(g,\omega)=\sum_i \frac{\lambda_i \langle \rho_k^{eq}[g-\langle g \rangle], \phi_i(g)\rangle}{j\omega-\lambda_i}\phi_i(g).
\label{eq46}
\end{equation}

This eigenvalue approach allows for very fast calculations of power spectral density:
\begin{equation}
\hat{S}_g(\omega)=2Re\left\{\sum_i \frac{\lambda_i}{j\omega-\lambda_i}\int_L g \langle \rho_k^{eq}[g-\langle g \rangle], \phi_i(g)\rangle\phi_i(g)dg\right\}.
\label{eq47}
\end{equation}

\section{Numerical Results}
The techniques described in the previous section have been numerically implemented for the case of uniaxial particles with and without the presence of the spin-torque effect. Some sample numerical results are presented below in figures 1 through 9 for the same effective anisotropy coefficient $K_{eff}$ and applied magnetic field $H_{az}$ but different parameters of the thermal noise.

\begin{figure}[!h]
\centering
\includegraphics[width=3.1in]{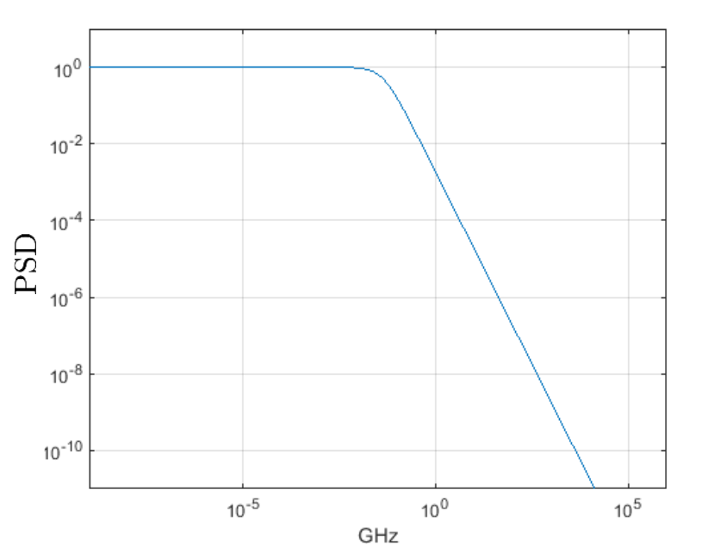}
\caption{Uniaxial particle with $K_{eff}=0.5, H_{az}=-0.7, \sigma^2=0.001, A=10^{12}$.}
\label{fig1}
\end{figure}

\begin{figure}[!h]
\centering
\includegraphics[width=3.1in]{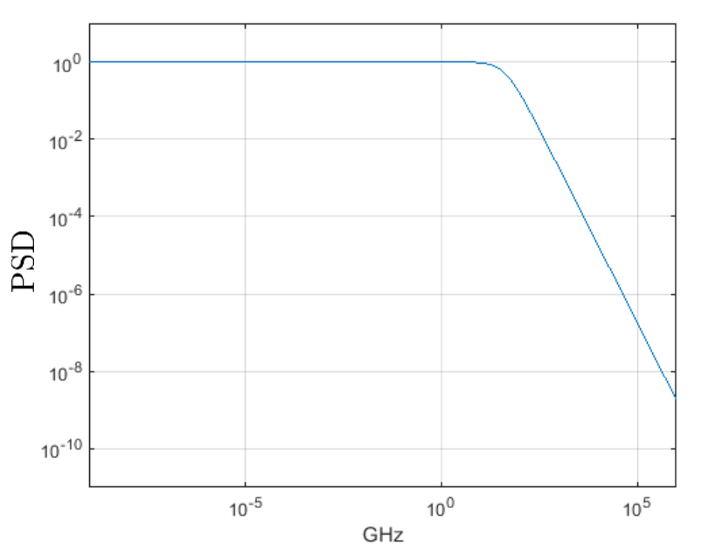}
\caption{Uniaxial particle with $K_{eff}=0.5, H_{az}=-0.7, \sigma^2=0.001, A=10^{15}$.}
\label{fig2}
\end{figure}

\begin{figure}[!h]
\centering
\includegraphics[width=3.1in]{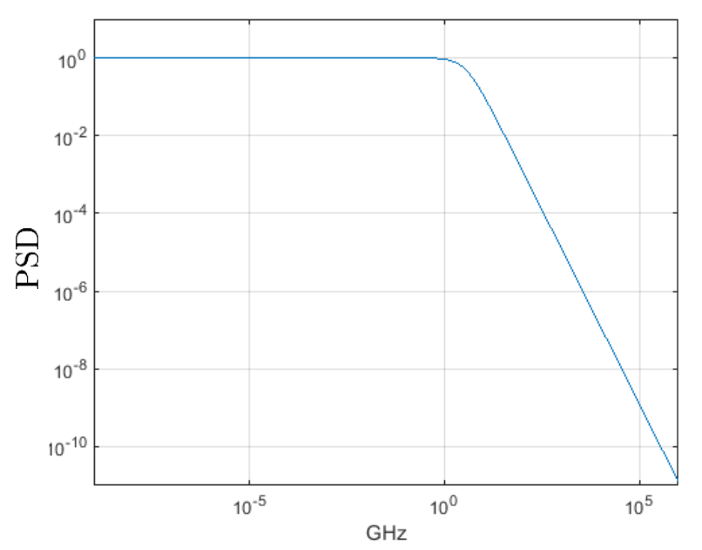}
\caption{Uniaxial particle with $K_{eff}=0.5, H_{az}=-0.7, \sigma^2=0.01, A=10^{12}$.}
\label{fig3}
\end{figure}

As seen in figures 1, 2, and 3, the power spectral density has a flat intensity response at lower frequencies before reaching a `knee' and decreasing as the frequency increases.  The placement of the `knee' is controlled by the strength of thermal noise and the roll-off appears to follow a $1/f^2$ dependence.

Indeed, by comparing figures 1 and 2 where the intensity of the noise is increased from $A=10^{12}$ to $A=10^{15}$, the placement of the knee is seen to have shifted to higher frequencies.  This is due to the fact that higher intensities of thermal noise cause more frequent jumps in magnetization.  Likewise, by comparing figures 1 and 3 where the variance of the noise is increased from $\sigma^2=0.001$ to $\sigma^2=0.01$, the placement of the knee also shifts to higher frequencies.  This is due to larger $\sigma^2$ allowing for larger jumps in magnetization and therefore an overall larger transition probability rate and scattering rate.  This reasoning is consistent with formula (\ref{eq6}).

\begin{figure}[!h]
\centering
\includegraphics[width=3.1in]{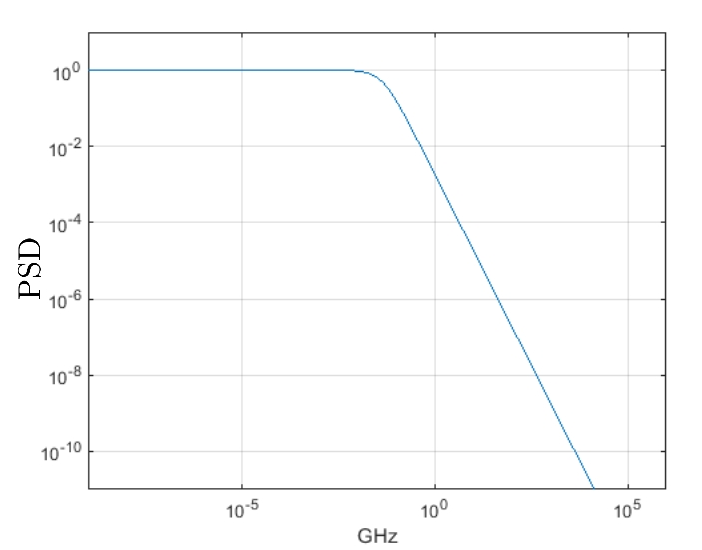}
\caption{PSD using the eigenvalue approach for an uniaxial particle with $K_{eff}=0.5, H_{az}=-0.7, \sigma^2=0.001, A=10^{12}$.}
\label{fig4}
\end{figure}

\begin{figure}[!h]
\centering
\includegraphics[width=3.1in]{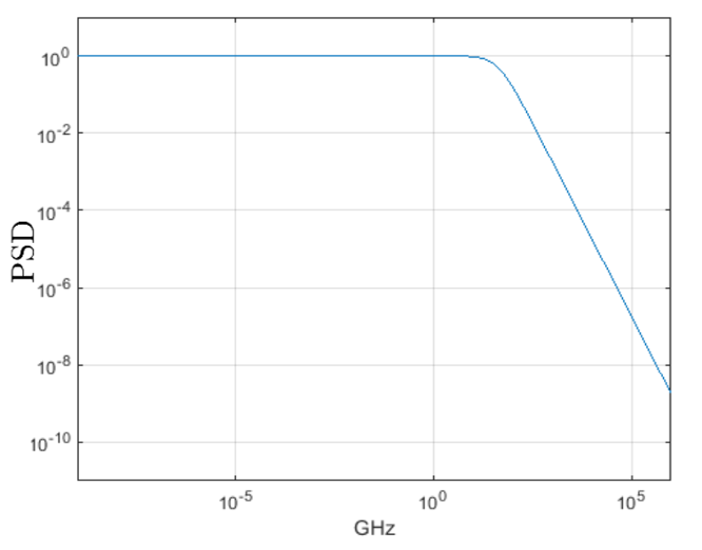}
\caption{PSD using the eigenvalue approach for an uniaxial particle with $K_{eff}=0.5, H_{az}=-0.7, \sigma^2=0.001, A=10^{15}$.}
\label{fig5}
\end{figure}

\begin{figure}[!h]
\centering
\includegraphics[width=3.1in]{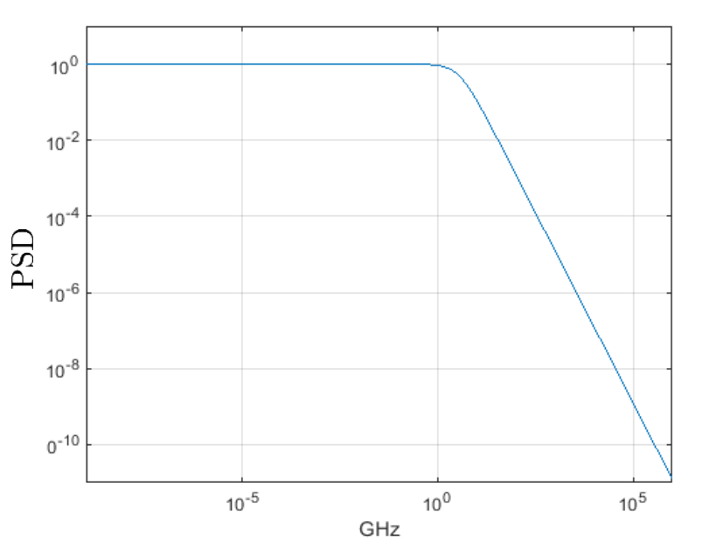}
\caption{PSD using the eigenvalue approach for an uniaxial particle with $K_{eff}=0.5, H_{az}=-0.7, \sigma^2=0.01, A=10^{12}$.}
\label{fig6}
\end{figure}

Figures 4, 5, and 6 represent the same power spectral densities as in Figures 1, 2, and 3 but calculated using the eigenvalue approach.  As seen from these figures, the power spectral densities exactly match their respective plots in Figures 1, 2, and 3.

\begin{figure}[!h]
\centering
\includegraphics[width=3.1in]{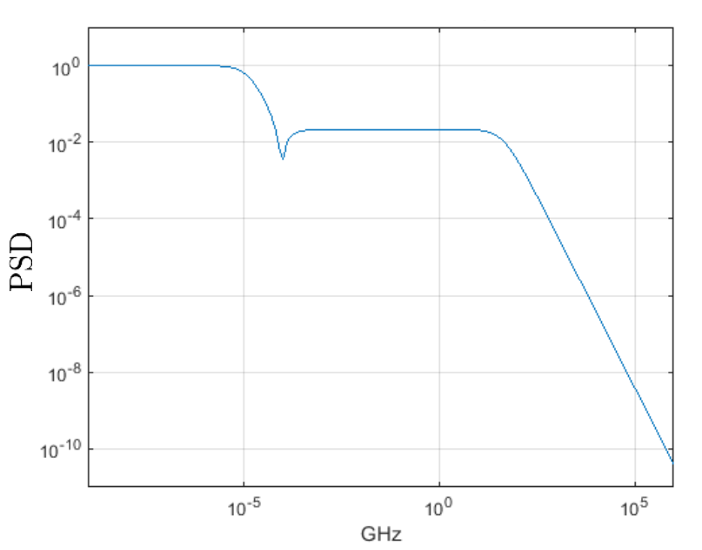}
\caption{Uniaxial particle with $K_{eff}=0.5, H_{az}=-0.7, \sigma^2=0.001, A=10^{15}$ and spin-torque characterized by $e_{pz}=-1, c_p=0.5, \beta=10^7$.}
\label{fig7}
\end{figure}

\begin{figure}[!h]
\centering
\includegraphics[width=3.1in]{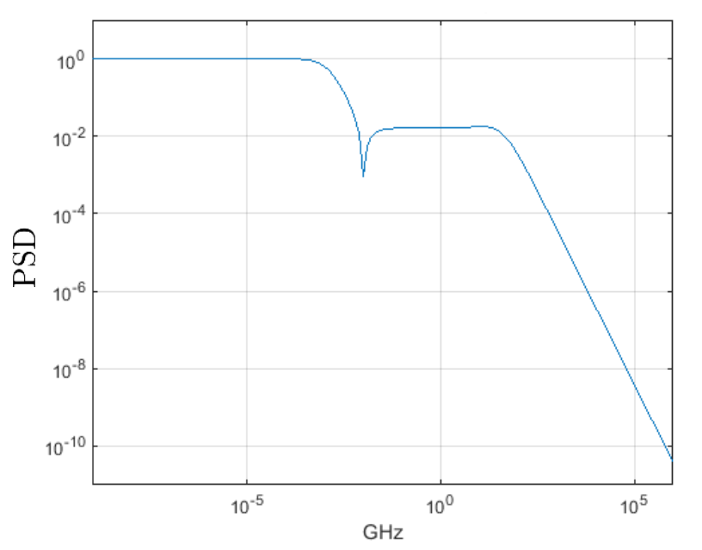}
\caption{Uniaxial particle with $K_{eff}=0.5, H_{az}=-0.7, \sigma^2=0.001, A=10^{15}$ and spin-torque characterized by $e_{pz}=-1, c_p=0.5, \beta=10^9$.}
\label{fig8}
\end{figure}

\begin{figure}[!h]
\centering
\includegraphics[width=3.1in]{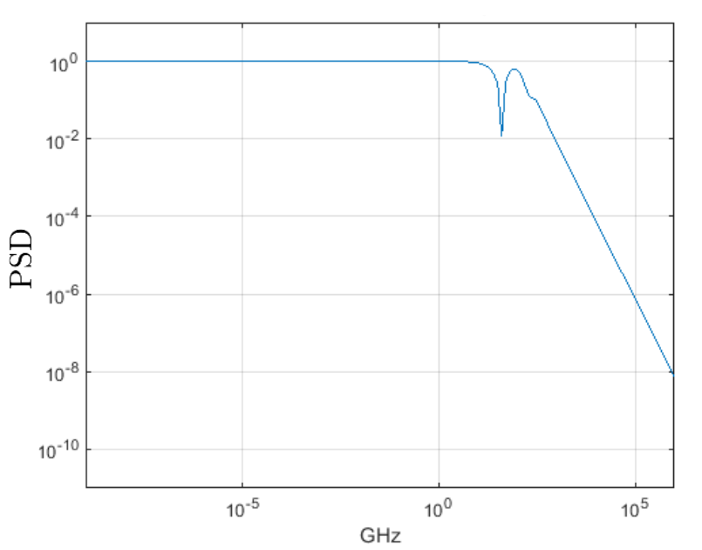}
\caption{Uniaxial particle with $K_{eff}=0.5, H_{az}=-0.7, \sigma^2=0.001, A=10^{15}$ and spin-torque characterized by $e_{pz}=-1, c_p=0.5, \beta=10^12$.}
\label{fig9}
\end{figure}

Figures 7, 8, and 9 show the effect of spin-torque on the power spectral density. The power spectral density is now seen to have two separate `knees'.  By comparing figure 7 with the corresponding power spectral density in the case of no spin-torque, we clearly see that the additional knee is a spin-torque effect.  This is also evident by comparing Figures 7, 8, and 9 where the strength of the spin torque ($\beta$) is increased from $\beta=10^7$ to $\beta=10^9$ and $\beta=10^{12}$.  The change in strength of the spin-torque shifts the location of the knee.

\section*{Acknowledgments}
This work was partially supported by Progetto Premiale MIUR-INRIM Nanotecnologie per la metrologia elettromagnetica, by MIUR-PRIN Project No. 2010ECA8P3 DyNanoMag, and by NSF.


\ifCLASSOPTIONcaptionsoff
  \newpage
\fi

\end{document}